# *In-vivo* high-resolution $\chi$-separation at 7T

Jiye Kim[1], Minjun Kim[1], Sooyeon Ji[1,2], Kyeongseon Min[1], Hwihun Jeong[1], Hyeong-Geol Shin[1,3,4], Chungseok Oh[1], Sina Straub[5], Seong-Gi Kim[6,7], and Jongho Lee[1,*]

[1]Laboratory for Imaging Science and Technology, Department of Electrical and Computer Engineering, Seoul National University, Seoul, Republic of Korea

[2]Division of Computer Engineering, Hankuk University of Foreign Studies, Yongin, Republic of Korea

[3]Department of Radiology, Johns Hopkins University School of Medicine, Baltimore, MD, USA

[4]F.M. Kirby Research Center for Functional Brain Imaging, Kennedy Krieger Institute, Baltimore, MD, USA

[5]Department of Radiology, Mayo Clinic, Jacksonville, FL, USA

[6]Center for Neuroscience Imaging Research, Institute for Basic Science, Suwon, Republic of Korea

[7]Department of Biomedical Engineering, Sungkyunkwan University, Suwon, Republic of Korea.

*Correspondence: jonghoyi@snu.ac.kr

## Abstract

A recently introduced quantitative susceptibility mapping (QSM) technique, $\chi$-separation, offers the capability to separate paramagnetic ($\chi_{para}$) and diamagnetic ($\chi_{dia}$) susceptibility distribution within the brain. *In-vivo* high-resolution mapping of iron and myelin distribution, estimated by $\chi$-separation, could provide a deeper understanding of brain substructures, assisting the investigation of their functions and alterations. This can be achieved using 7T MRI, which benefits from a high signal-to-noise ratio and susceptibility effects. However, applying $\chi$-separation at 7T presents difficulties due to the requirement of an $R_2$ map, coupled with issues such as high specific absorption rate (SAR), large $B_1$ transmit field inhomogeneities, and prolonged scan time. To address these challenges, we developed a novel deep neural network, R2PRIMEnet$_{7T}$, designed to convert a 7T $R_2^*$ map into a 3T $R_2'$ map. Building on this development, we present a new pipeline for $\chi$-separation at 7T, enabling us to generate high-resolution $\chi$-separation maps from multi-echo gradient-echo data. The proposed method is compared with alternative pipelines, such as an end-to-end network and linearly-scaled $R_2'$, and is validated against $\chi$-separation maps at 3T, demonstrating its accuracy. The 7T $\chi$-separation maps generated by the proposed method exhibit similar contrasts to those from 3T, while 7T high-resolution maps offer enhanced clarity and detail. Quantitative analysis confirms that the proposed method surpasses the alternative pipelines. The proposed method results well delineate the detailed brain structures associated with iron and myelin. This new pipeline holds promise for analyzing iron and myelin concentration changes in various neurodegenerative diseases through precise structural examination.

## 1. Introduction

Iron and myelin play pivotal roles in maintaining normal brain function, with their dysregulation implicated in neurodegenerative diseases such as multiple sclerosis, Parkinson′s disease, and Alzheimer′s disease (Compston and Coles, 2008; Zecca et al., 2004). These substances hold promise as biomarkers for such diseases, emphasizing the need for *in-vivo* imaging methods to map the distribution of iron and myelin.

Quantitative susceptibility mapping (QSM) has emerged as a powerful neuroimaging technique using MRI to quantify tissue magnetic susceptibility (Rochefort et al., 2008; Shmueli et al., 2009). Previous studies have demonstrated the sensitivity of QSM in detecting susceptibility sources related to brain disorders such as multiple sclerosis, and Parkinson′s disease (Langkammer et al., 2013; Lotfipour et al., 2012). However, since iron is paramagnetic and myelin is diamagnetic, QSM faces challenges in quantifying their contributions due to its limitation of presenting cumulative susceptibility. This has led to a growing need to measure the susceptibility contributions of iron and myelin separately. A recent advancement in QSM, $\chi$-separation (Shin et al., 2021), is a newly proposed QSM method that can disentangle paramagnetic ($\chi_{para}$) and diamagnetic ($\chi_{dia}$) susceptibility distribution in the brain (Chen et al., 2021; Dimov et al., 2022; Emmerich et al., 2021a, 2021b). $\chi$-separation jointly employs an $R_2′$ (= $R_2^*$- $R_2$) map and a local field map to generate $\chi_{para}$ and $\chi_{dia}$ maps. Studies have demonstrated the correspondence between paramagnetic susceptibility and iron concentration and between diamagnetic susceptibility and myelin concentration in the brain (Lee et al., 2024; Li et al., 2023; Min et al., 2024; Shin et al., 2021).

In recent years, 7T imaging has become a popular option for high-resolution images, primarily due to its high signal-to-noise ratio (SNR), which enhances the image quality of high-resolution MRI (Deistung et al., 2013b; Duyn et al., 2007; Straub et al., 2019). In QSM, the benefits of 7T are further enhanced by the susceptibility effect, which increases proportionally with the B0 field strength. We expect the same benefits in $\chi$-separation, potentially providing a sub-millimeter resolution map for brain structures. However, applying $\chi$-separation at 7T poses challenges due to the requirement of an $R_2$ map in $\chi$-separation. This $R_2$ map is generated from a spin-echo sequence with multiple echo times, suffering from a high specific absorption rate (SAR), severe $B_1$ transmit field inhomogeneities, and long scan time. Efforts have been made for $R_2$ mapping at 7T (Cox and Gowland, 2010; Emmerich et al., 2019) but practical approaches for whole-brain high-resolution $R_2$ mapping remain as a challenge. Recently, a deep neural network was proposed to eliminate the necessity for the $R_2$ map in $\chi$-separation (M. Kim et al., 2024). This development aligns with a growing trend where deep learning approaches have demonstrated successful adoption in MRI reconstruction, particularly in tackling ill-posed problems such as QSM (Yoon et al., 2018), fast imaging (Hammernik et al., 2018; Kwon et al., 2017; Lee et al., 2017), and contrast synthesis (Hagiwara et al., 2019; Moya-Sáez et al., 2021; Tanenbaum et al., 2023). However, the network for $\chi$-separation (M. Kim et al., 2024) is designed to work for 3T $R_2^*$ input and, therefore, cannot be directly applicable to 7T $R_2^*$ input because of the field-dependent $R_2^*$. Consequently, further research is required to generate a $\chi$-separation map at 7T.

This study aims to produce *in-vivo* high-resolution $\chi$-separation maps at 7T. To achieve this goal, we introduce a novel deep neural network, R2PRIMEnet$_{7T}$, designed to convert 7T $R_2^*$ to 3T $R_2'$. Then, a new data processing pipeline that includes the network is developed to process high-resolution 7T multi-echo gradient-echo data for $\chi$-separation. The resulting 7T $\chi$-separation maps are validated with respect to 3T $\chi$-separation maps. We also conduct a comparative analysis with alternative methods, including a pipeline using an end-to-end network and a linearly-scaled $R_2'$ input. The proposed method is applied to delineate detailed brain structures, demonstrating the benefits of the unique contrasts of myelin and iron as well as increased resolution of 0.65 mm isotropic resolution at 7T.

## 2. Methods

*2.1. MRI data acquisition and processing*

Ten healthy volunteers (mean age: 26 ± 2.3 years old; 3 females and 7 males) were scanned at both 3T and 7T MRI (Siemens Tim Trio for 3T and Magnetom Terra for 7T, Erlangen, Germany). All volunteers signed a written consent form approved by the institutional review board. The sequences and their acquisition parameters are listed in Table 1. At 3T, 1 mm isotropic resolution multi-echo gradient-echo (GRE) was acquired at six different head orientations for $R_2^*$ and local field maps. Additionally, a 2D multi-echo spin-echo (SE) sequence was utilized for an $R_2$ map, and a 3D magnetization-prepared rapid gradient echo (MPRAGE) sequence was acquired for a $T_1$-weighted image. At 7T, single-orientation 3D multi-echo GRE data for high-resolution $R_2^*$ and local field maps was acquired in the same volunteers as 3T. The resolutions for 7T were 0.65 mm isotropic resolution in eight out of the ten subjects (GRE #1), $0.70 \times 0.70 \times 0.75$ mm$^3$ resolution in one subject (GRE #2), and 0.60 mm isotropic resolution in one subject (GRE #3). The whole-brain $R_2$ map at 7T could not be acquired due to high SAR and long scan time issues of multi-echo SE sequences.

**Table 1.** MRI acquisition parameters at 3T and 7T.

| MRI acquisition parameters | 3T GRE | 3T SE | 3T MPRAGE | 7T GRE #1 | 7T GRE #2 | 7T GRE #3 |
|---|---|---|---|---|---|---|
| TR (ms) | 38 | 7800 | 2400 | 32 | 31 | 38 |
| TEs (ms) | 7.7, 12.7, 17.8, 22.8, 27.8, 32.9 | 15, 30, 45, 60, 75, 90 | 2.1 | 4.5, 9.0, 13.5, 18.0, 22.5 | 4.5, 9.0, 13.5, 18.0, 22.5 | 9.3, 18.0, 26.7 |
| FOV (mm$^3$ or mm$^2$) | 256 × 224 × 160 | 256 × 204 | 256 × 256 × 224 | 228 × 185 × 146 | 210 × 204 × 144 | 227 × 185 × 105 |
| Voxel size (mm$^3$ or mm$^2$) | 1.0 × 1.0 × 1.0 | 1.0 × 1.0 | 1.0 × 1.0 × 1.0 | 0.65 × 0.65 × 0.65 | 0.70 × 0.70 × 0.75 | 0.60 × 0.60 × 0.60 |
| Slice thickness (mm) | | 2.0 | | | | |
| Number of slices | | 76 | | | | |
| Bandwidth (Hz/pixel) | 290 | 110 | 210 | 250 | 320 | 140 |
| Total acquisition time | 6 min 42 sec | 13 min 25 sec | 5 min 45 sec | 8 min 23 sec | 6 min 30 sec | 21 min 11 sec |
| Number of subjects | 10 subjects | 10 subjects | 10 subjects | 8 subjects | 1 subject | 1 subject |

The data processing for the 3T dataset was carried out as follows: A brain mask was extracted from the sum of squares of the multi-echo GRE magnitude images using the brain extraction tool algorithm in FSL (Jenkinson et al., 2012). $R_2^*$ mapping was performed by fitting a mono-exponential decay function to the GRE magnitude images using a fast mono-exponential fitting algorithm (Pei et al., 2015). The GRE phase data underwent spatial unwrapping (Dymerska et al., 2021), weighted echo averaging (Wu et al., 2012a), and background-field removal (Schweser et al., 2011; Wu et al., 2012b), producing a local field map. The magnitude images of the first echo from the five rotated head orientations were individually registered to that of the unrotated head orientation using FLIRT in FSL (Jenkinson et al., 2012). Then, the resulting rotation matrix was applied to the local field map and $R_2^*$ map for registration. Using the registered local field maps and rotation information, a QSM map was generated via the COSMOS algorithm (Liu et al., 2009). For $R_2$ mapping, the multi-echo SE magnitude signal from each voxel was matched to a computer-simulated dictionary of spin-echo decay curves incorporating stimulated echoes. This dictionary was constructed with the StimFit toolbox (https://github.com/rmlebel/StimFit), employing a slice profile from an actual RF pulse. This $R_2$ map was then registered to the six $R_2^*$ maps of each orientation using FLIRT. An $R_2'$ map was calculated through a voxel-wise subtraction of the registered $R_2$ map from the $R_2^*$ map of each direction, with negative $R_2'$ values set to zero to enforce physics. These $R_2'$ maps from each orientation were subsequently registered to the unrotated head orientation using the same rotation matrix applied for the local field and $R_2^*$ maps. Finally, the paramagnetic ($\chi_{para}$) and diamagnetic ($\chi_{dia}$) susceptibility maps were generated by the multi-orientation $\chi$-separation reconstruction algorithm ($\chi$-sep-COSMOS) (Shin et al., 2022).

For data processing in 7T, the GRE data underwent the same processing steps as in the 3T GRE data to generate $R_2^*$ and local field maps. These steps included brain mask extraction, mono-exponential fitting of the GRE magnitude images for $R_2^*$ mapping, spatial unwrapping, weighted echo averaging, and background field removal of the GRE phase images. While the 3T data processing involved multi-orientation processing, only single-orientation processing was applied to the 7T data.

## 2.2. χ-separation pipelines at 7T

The overall data processing pipelines for high-resolution $\chi$-separation using 7T data are illustrated in Figure 1. We developed and evaluated three pipelines, each designed to circumvent the requirement for an $R_2$ map: (a) $\chi$-separation using an R2PRIMEnet$_{7T}$ (Proposed; Figure 1a), (b) $\chi$-separation using an end-to-end network (Alternative 1; Figure 1b), and (c) $\chi$-separation using linearly-scaled $R_2'$ (Alternative 2; Figure 1c).

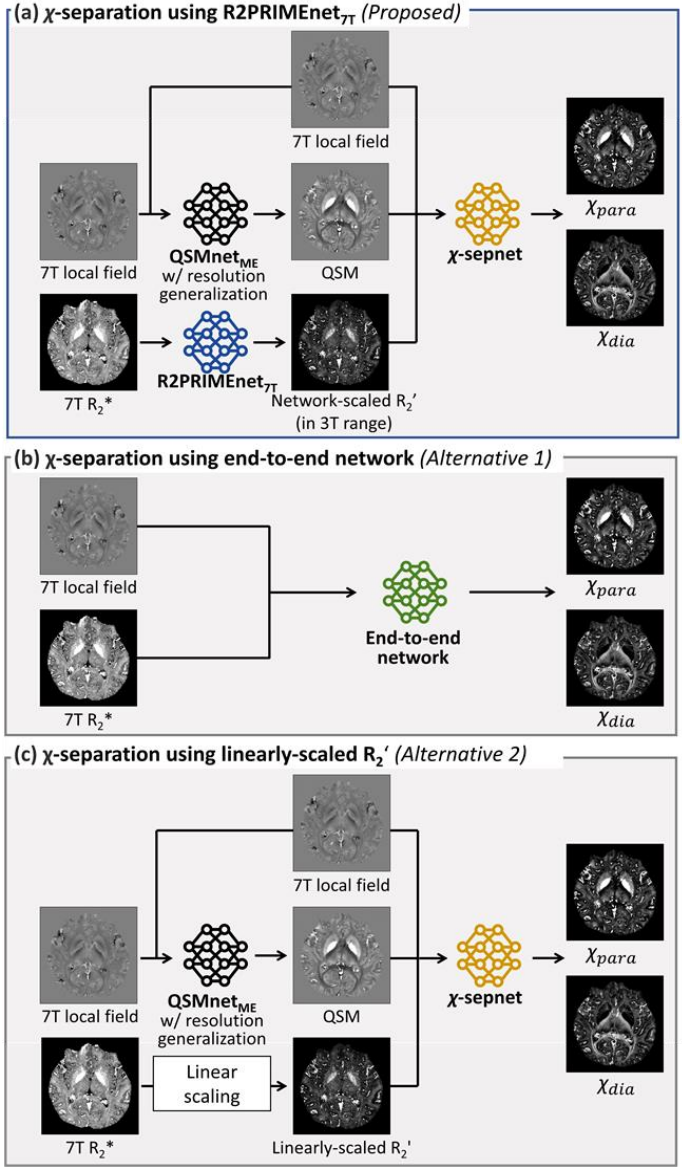

**Figure 1.** Overview of $\chi$-separation pipelines at 7T. (a) Proposed $\chi$-separation pipeline at 7T utilizes a deep neural network, R2PRIMEnet$_{7T}$, to address the challenges of acquiring a whole-brain high-resolution $R_2$ map at 7T. A high-resolution QSM map is reconstructed from a 7T local field map using QSMnet$_{ME}$, with a resolution generalization technique. A 7T $R_2^*$ map is converted to a 3T $R_2'$ map using a newly developed R2PRIMEnet$_{7T}$. The local field, QSM, and network-scaled $R_2'$ maps are then fed into $\chi$-sepnet, resulting in high-resolution $\chi$-separation maps. (b) An alternative pipeline uses an end-to-end network designed to predict $\chi$-separation maps directly from 7T local field and $R_2^*$ maps. (c) Another alternative pipeline uses a linearly-scaled $R_2'$ map as the $R_2'$ input for $\chi$-sepnet.

*2.2.1. Proposed pipeline: χ-separation using R2PRIMEnet$_{7T}$*

The proposed pipeline employed 3T χ-sepnet (M. Kim et al., 2024), which takes the three maps of local field, QSM, and 3T $R_2'$ as the input and produces COSMOS-quality χ-separation maps as the output. For 7T χ-separation processing, a 7T local field map and a QSM map generated by QSMnet$_{ME}$ with resolution generalization (see Supplementary Information for details) were applied as the input for 3T χ-sepnet because QSM is field-independent (Spincemaille et al., 2020). For the $R_2'$ input of χ-sepnet, a 7T $R_2^*$ map was converted to a 3T $R_2'$ map via a newly-designed network, R2PRIMEnet$_{7T}$, and then was applied (Fig. 1a).

For the development and evaluation of R2PRIMEnet$_{7T}$, the ten healthy volunteers' $R_2'$ maps from 3T and $R_2^*$ maps from 7T were used (five for training, one for validation, and four for test). The 7T $R_2^*$ maps were under-sampled by truncating k-space to match the resolution of the 3T $R_2'$ maps and then registered to the 3T $R_2'$ maps using the symmetric diffeomorphic image normalization algorithm in Advanced Normalization Tools (ANTs) (Avants et al., 2009). To increase the size of the training dataset, rotation augmentation was applied using six randomly generated rotation matrices, which were the same for every subject. Considering fiber orientation-dependent $R_2^*$ in white matter (Lee et al., 2011; Oh et al., 2013), the random rotation was confined to the plane perpendicular to the B0 direction. The training dataset was divided into 4200 patches of 64 × 64 × 64 voxels, with each map split into 6 patches along the x-axis, 5 along the y-axis, and 4 along the z-axis. The network followed a 3D U-net architecture (see Supplementary Information and Figure S1 for details) (Yoon et al., 2018). The neural network was trained to take a patch of the 7T $R_2^*$ maps as the input and the corresponding patch of the 3T $R_2'$ maps as the label. The loss function ($loss_{R2PRIME_{7T}}$) included two voxel-wise differences ($loss_{L1}$: L1 loss, and $loss_{RMSLE}$: root mean squared log error loss) and one edge preservation ($loss_{Gradient}$: gradient loss):

$$loss_{L1} = \left\| R_{2\,3T}' - f(R_{2\,7T}^*) \right\|_1, \qquad (1)$$

$$loss_{RMSLE} = \left\| \log(R_{2\,3T}' + 1) - \log(f(R_{2\,7T}^*) + 1) \right\|_2, \qquad (2)$$

$$loss_{Gradient} = \left\| |\nabla R_{2\,3T}'|_x - |\nabla f(R_{2\,7T}^*)|_x \right\|_1 + \left\| |\nabla R_{2\,3T}'|_y - |\nabla f(R_{2\,7T}^*)|_y \right\|_1 + \left\| |\nabla R_{2\,3T}'|_z - |\nabla f(R_{2\,7T}^*)|_z \right\|_1, \qquad (3)$$

$$loss_{R2PRIME_{7T}} = loss_{L1} + +w_1 loss_{RMSLE} + w_2 loss_{Gradient}, \qquad (4)$$

where $R_2'_{3T}$ is the label R₂' from 3T, $R_2^*_{7T}$ is the input R₂* from 7T, $f$ denotes R2PRIMEnet₇ₜ, and the weights of $w_1 = 1.0$ and $w_2 = 0.1$. During the training, a batch size of 12 was used with an Adam optimizer and a learning rate of 0.002. The training process ran for 50 epochs, and the epoch with the best validation normalized root-mean-square-error (NRMSE) was selected.

*2.2.2. Alternative 1: χ-separation using end-to-end network*

To evaluate the performance of the proposed pipeline, we investigated an alternative approach using an end-to-end network. The end-to-end network was designed to directly predict χ-separation maps from 7T local field and R₂* maps (Fig. 1b).

For the end-to-end network, the 3D U-net architecture utilized in χ-sepnet was modified for two inputs (i.e., local field and R₂* maps) instead of three (i.e., local field, QSM, and R₂* maps) and deployed. The same datasets as in the proposed method were utilized and the preprocessing including resolution matching and registration for the datasets were the same. The training dataset consisted of 7T local field and R₂* maps as the input and the 3T χ-sep-COSMOS maps as the label. Rotation augmentation was applied using the same augmentation method as in R2PRIMEnet₇ₜ. The datasets were segmented into patches of 64 × 64 × 64 voxels, creating a total of 11760 patches, with each map split into 8 patches along the x-axis, 7 along the y-axis, and 6 along the z-axis.

During training, a batch size of 12 was utilized with RMSprop optimizer, setting the learning rate at 0.003. The employed loss function ($\mathcal{L}_{end}$) included voxel-wise difference ($\mathcal{L}_{L1}$), edge preservation loss ($\mathcal{L}_{Gradient}$), and model loss ($\mathcal{L}_{Model}$). These training parameters were adopted from χ-sepnet (M. Kim et al., 2024) with the note that the QSM model loss ($loss_{model}^{QSM}$) was excluded, because this network does not take a QSM map as input. The employed loss function is outlined below:

$$\mathcal{L}_{end} = \mathcal{L}_{L1} + h_1 \mathcal{L}_{Gradient} + h_2 \mathcal{L}_{Model}$$

$$\mathcal{L}_{L1} = \left\| g(\varphi, R_2^*_{7T})_{para} - \chi_{para} \right\|_1 + \left\| g(\varphi, R_2^*_{7T})_{dia} - \chi_{dia} \right\|_{1,}$$

$$\mathcal{L}_{Gradient} = \left\| \left| \nabla g(\varphi, R_{2\,7T}^*)_{para} \right|_x - |\nabla \chi_{para}|_x \right\|_1 + \left\| \left| \nabla g(\varphi, R_{2\,7T}^*)_{dia} \right|_x - |\nabla \chi_{dia}|_x \right\|_1$$

$$+ \left\| \left| \nabla g(\varphi, R_{2\,7T}^*)_{para} \right|_y - |\nabla \chi_{para}|_y \right\|_1 + \left\| \left| \nabla g(\varphi, R_{2\,7T}^*)_{dia} \right|_y - |\nabla \chi_{dia}|_y \right\|_1$$

$$+ \left\| \left| \nabla g(\varphi, R_{2\,7T}^*)_{para} \right|_z - |\nabla \chi_{para}|_z \right\|_1 + \left\| \left| \nabla g(\varphi, R_{2\,7T}^*)_{dia} \right|_z - |\nabla \chi_{dia}|_z \right\|_1,$$

$$\mathcal{L}_{Model} = \mathcal{L}_{Model}^{field} + \mathcal{L}_{Model}^{R_2'},$$

$$\mathcal{L}_{Model}^{field} = \left\| \left( d * \left( g(\varphi, R_{2\,7T}^*)_{para} - \left| g(\varphi, R_{2\,7T}^*)_{dia} \right| \right) \right) - \varphi \right\|_1,$$

$$\mathcal{L}_{Model}^{R_2'} = \left\| D_r \cdot \left( g(\varphi, R_{2\,7T}^*)_{para} + \left| g(\varphi, R_{2\,7T}^*)_{dia} \right| \right) - R_2' \right\|_1,$$

where $g$ denotes the end-to-end network, $h_1 = 0.1$, and $h_2 = 1$, $\varphi$, $R_{2\,7T}^*$, and $R_2'$ refer to the local field, 7T $R_2^*$, and 3T $R_2'$ maps, respectively. $\chi_{para}$ and $\chi_{dia}$ are the labels that correspond to para- and diamagnetic susceptibility maps reconstructed by $\chi$-sep-COSMOS, $d$, $*$, and $D_r$ (=114 Hz/ppm) refer to dipole kernel, convolution operation, and relaxometric constant, respectively.

The training process ran for 60 epochs, and the network was selected based on the best validation NRMSE. For inference, the 7T local field and $R_2^*$ maps are fed into the network (Figure 1b).

*2.2.3. Alternative 2: χ-separation using linearly-scaled $R_2'$*

This approach investigated the feasibility of linearly-scaling 7T $R_2^*$ to 3T $R_2'$. It is motivated by a recent study in which 3T $R_2^*$ was linearly-scaled to substitute $R_2'$ (Dimov et al., 2022; Ji et al., 2024). To determine the linear scaling factor between 3T $R_2'$ and 7T $R_2^*$, regions of interest (ROIs) from the χ-separation atlas (Min et al., 2024) were utilized and the mean ROI values from the five training subjects were calculated. The ROIs for each subject were segmented as follows (Min et al., 2024): a QSM map was linearly combined with an intensity-normalized T1-weighted image to create a hybrid image. This hybrid image was then registered to the hybrid image atlas in the MNI space from MuSus-100 (He et al., 2023) via the symmetric diffeomorphic image normalization algorithm in ANTs. Subsequently, the ROIs in the χ-separation atlas (Table 2) were inverse-transformed from the MNI space to each subject's space. Linear regression of the ROI-averaged values at 3T and 7T was conducted, and the slope of the regression line, which was 0.190 (see Figure S2), was used as the scaling factor. Finally, a linearly-scaled $R_2'$ map served as $R_2'$ input in χ-sepnet along with local field and QSM maps, which were processed by the same approach as in the proposed method (Figure 1c).

**Table 2**. Regions of interest (ROI) utilized for the ROI analysis. The ROIs include subcortical nuclei, thalamic nuclei, and white matter.

|  | ROI name |  | ROI name |
|---|---|---|---|
| Subcortical nuclei | Caudate | White matter | Genu of corpus callosum |
|  | Putamen |  | Body of corpus callosum |
|  | Globus pallidus |  | Splenium of corpus callosum |
|  | Nucleus accumbens |  | Cerebral peduncle |
|  | Substantia nigra |  | Anterior limb of internal capsule |
|  | Red nucleus |  | Posterior limb of internal capsule |
|  | Ventral pallidum |  | Retrolenticular part of internal capsule |
|  | Subthalamic nucleus |  | Anterior corona radiata |
| Thalamus | Medial thalamic nuclei |  | Superior corona radiata |
|  | Lateral thalamic nuclei |  | Posterior corona radiata |
|  | Pulvinar |  | Posterior thalamic radiation |
|  |  |  | Sagittal stratum |
|  |  |  | Superior longitudinal fasciculus |

*2.3. Evaluation*

To assess the performance of R2PRIMEnet$_{7T}$, we compared network-scaled $R_2'$ and linearly-scaled $R_2'$ maps against the 3T $R_2'$ map (reference) in the four test subjects. A quantitative comparison was conducted using peak signal-to-noise ratio (pSNR), normalized root-mean-square-error (NRMSE), high-frequency error norm (HFEN), and structural similarity index (SSIM) with respect to the reference. When computing these metrics, cerebrospinal fluid (CSF) (Liu et al., 2018) and blood vessels (T. Kim et al., 2024) were masked out since large vessels induce artifacts due to non-local effects and flow (Lee et al., 2024). The means and standard deviations of the metrics across the four test subjects were computed.

For 7T χ-separation, the three pipelines in Fig. 1 were evaluated using the four test subjects, whose GRE data were acquired at a 0.65 mm isotropic resolution. As a reference, 3T χ-separation maps were generated using χ-sepnet-$R_2^*$, a pipeline for generating χ-separation maps from 3T multi-echo GRE data (M. Kim et al., 2024). In this pipeline, a 3T $R_2^*$ map is converted into 3T $R_2'$ map using R2PRIMEnet, and χ-sepnet then generates χ-separation maps from converted $R_2'$ map, local field map, and QSM map produced by QSMnet$_{ME}$ (see Supplementary Information for details). Additionally, a comparison with respect to 3T χ-sep-COSMOS maps was also conducted. All test data were processed in the original resolution (0.65 mm isotropic voxel). Then, the χ-separation maps were under-sampled to a 1 mm isotropic resolution by k-space truncation and registered to the 3T χ-separation maps using ANTs (symmetric diffeomorphic image normalization algorithm). The resulting maps were analyzed by pSNR, NRMSE, HFEN, and SSIM with respect to the reference, excluding CSF and vessels. In addition, an ROI analysis was performed, comparing the mean values of the ROIs (Table 2) between the 7T χ-separation map and the 3T reference map. The same ROIs as described in Section 2.2.3 were utilized (eleven subcortical nuclei and thalamic nuclei regions for $\chi_{para}$ and thirteen white matter fiber tracts for $\chi_{dia}$). The ROI analysis was conducted utilizing linear regression and the Bland-Altman plot.

*2.4. Applications*

To evaluate the quality and utility of the high-resolution χ-separation maps at 7T, qualitative analysis was performed focusing on its ability to delineate detailed brain structures. We compared χ-separation maps with 0.65 mm isotropic resolution against those from 3T with 1 mm isotropic resolution, examining matching locations.

All data processing was performed using MATLAB (MATLAB 2021a, MathWorks Inc., Natick, MA, USA). The deep neural networks are implemented using PyTorch (Paszke et al., 2019) in Nvidia RTX 8000 GPU (Nvidia Corp., Santa Clara, CA).

## 3. Results

### 3.1. R2PRIMEnet$_{7T}$

To demonstrate the efficacy of the proposed R2PRIMEnet$_{7T}$, we compare the 7T R$_2$* maps (input, first column), 3T R$_2'$ maps (label, second column), R2PRIMEnet$_{7T}$-generated R$_2'$ maps, (output, third column), and linearly-scaled R$_2'$ maps (last column) (Fig. 2). The 3T R$_2'$ maps and the R2PRIMEnet$_{7T}$ output maps exhibit similar contrasts, with the latter appearing less noisy. In the difference maps, noise-like error is observed. On the other hand, the linearly-scaled R$_2'$ maps reveal higher errors in the difference maps, particularly in the grey matter regions. The mean pSNR, NRMSE, HFEN, and SSIM report 28.9 ± 0.6 dB, 67.0 ± 2.9%, 76.5 ± 3.9%, and 0.882 ± 0.044, respectively for the network output maps and 27.6 ± 0.7 dB, 77.7 ± 4.0%, 91.9 ± 3.7%, and 0.860 ± 0.050, respectively for the linearly-scaled R$_2'$ maps, indicating better correspondence of the network-scaled R$_2'$ maps to the labels.

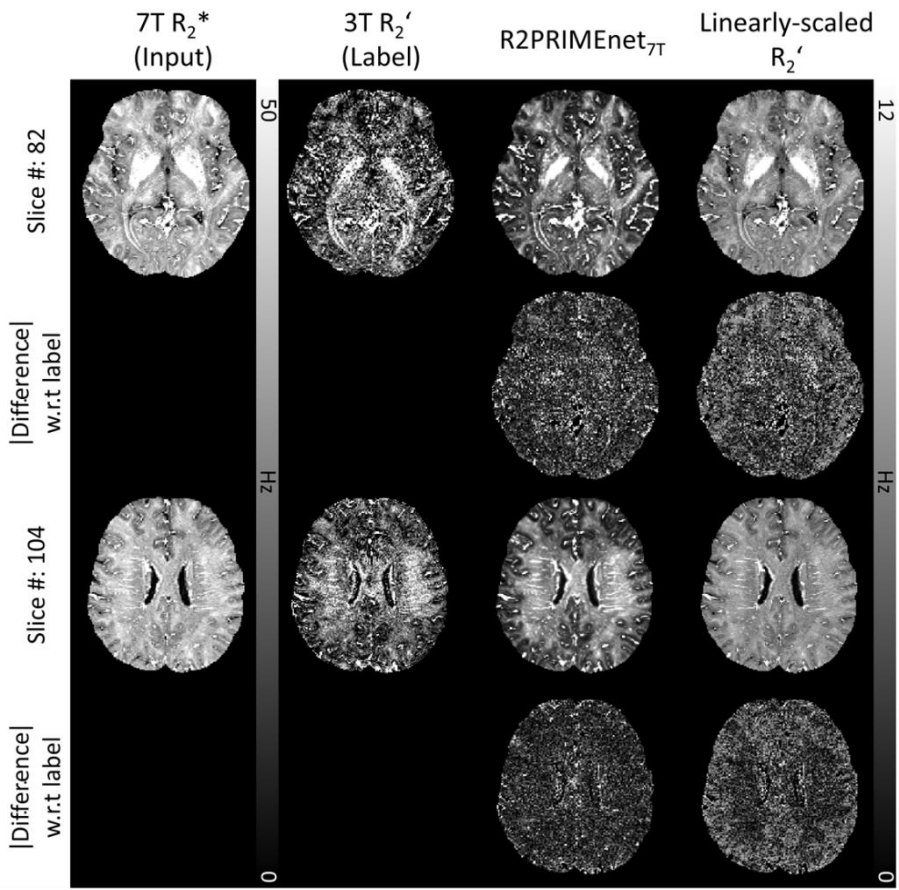

**Figure 2.** Results of R2PRIMEnet$_{7T}$. R$_2$* maps at 7T (input; first column), R$_2'$ maps at 3T (label; second column), output maps using R2PRIMEnet$_{7T}$ (output; third column), and linearly-

scaled R$_2'$ (last column) are displayed. In the difference maps, CSF and vessels are masked out. The 3T R$_2'$ maps and output maps of R2PRIMEnet$_{7T}$ show similar contrasts, while network output maps appear cleaner. Conversely, linearly-scaled R$_2'$ maps show more noticeable errors and different contrasts.

*3.2 Comparison of three χ-separation methods at 7T*

In Figure 3, χ-separation maps from 3T χ-sepnet-R$_2$* (reference, first column), proposed method (blue boxes, second column), end-to-end network (alternative 1, third column), and linearly-scaled R$_2'$ (alternative 2, last column) are displayed. For both $\chi_{para}$ and $\chi_{dia}$ maps, all three 7T methods achieve contrasts similar to those of 3T χ-sepnet-R$_2$*. Difference maps reveal that the proposed method produces overall smallest error. Table 3 summarizes the quantitative metrics for the three 7T methods. The metrics demonstrate that the proposed method reports the best correspondence to the reference for both $\chi_{para}$ and $\chi_{dia}$.

Figure 4 shows the scatter plots of the ROIs. Each dot depicts the mean ROI value of a subject from 3T χ-sepnet-R$_2$* on the x-axis and the corresponding value from each 7T χ-separation method on the y-axis (blue: proposed, red: end-to-end network, black: linearly-scaled R$_2'$). The linear regression results show the highest slope of 0.924 and $R^2$ of 0.995 for the proposed method, demonstrating close alignment to the 3T χ-sepnet-R$_2$* results. The linearly-scaled R$_2'$ yields a slightly lower slope of 0.919 ($R^2 = 0.994$) while the end-to-end network shows the lowest slope of 0.863 ($R^2 = 0.986$). These slopes, all of which are less than 1, suggest that all three methods slightly underestimate the susceptibility values compared to the results of 3T χ-sepnet-R$_2$* (see Discussion).

As shown in the Bland-Altman plots (Figure 5), all methods exhibit small biases (solid blue line; proposed method: 0.00463 ppm for $\chi_{para}$, 0.00511 ppm for $\chi_{dia}$, end-to-end network: 0.0116 ppm for $\chi_{para}$, 0.00326 ppm for $\chi_{dia}$, and linearly-scaled R$_2'$: 0.00448 ppm for $\chi_{para}$, 0.00470 ppm for $\chi_{dia}$). The proposed method displays the narrowest limits of agreement for both $\chi_{para}$ and $\chi_{dia}$, suggesting good agreement to the χ-separation at 3T (proposed method: 0.0248 for $\chi_{para}$, 0.0124 for $\chi_{dia}$, end-to-end network: 0.0445 for $\chi_{para}$, 0.0181 for $\chi_{dia}$, and linearly-scaled R$_2'$: 0.0295 for $\chi_{para}$, 0.0130 for $\chi_{dia}$).

When using 3T $\chi$-sep-COSMOS maps as a reference, (see Figure S3, S4, S5, and Table S1), we observed similar trends, with the proposed method demonstrating the closest alignment in contrast and lowest error..

All of these results including the difference maps in Figure 3, ROI analysis, and Bland-Altman plots, suggest that the proposed method yields high-quality 7T $\chi$-separation results that match the 3T outcomes.

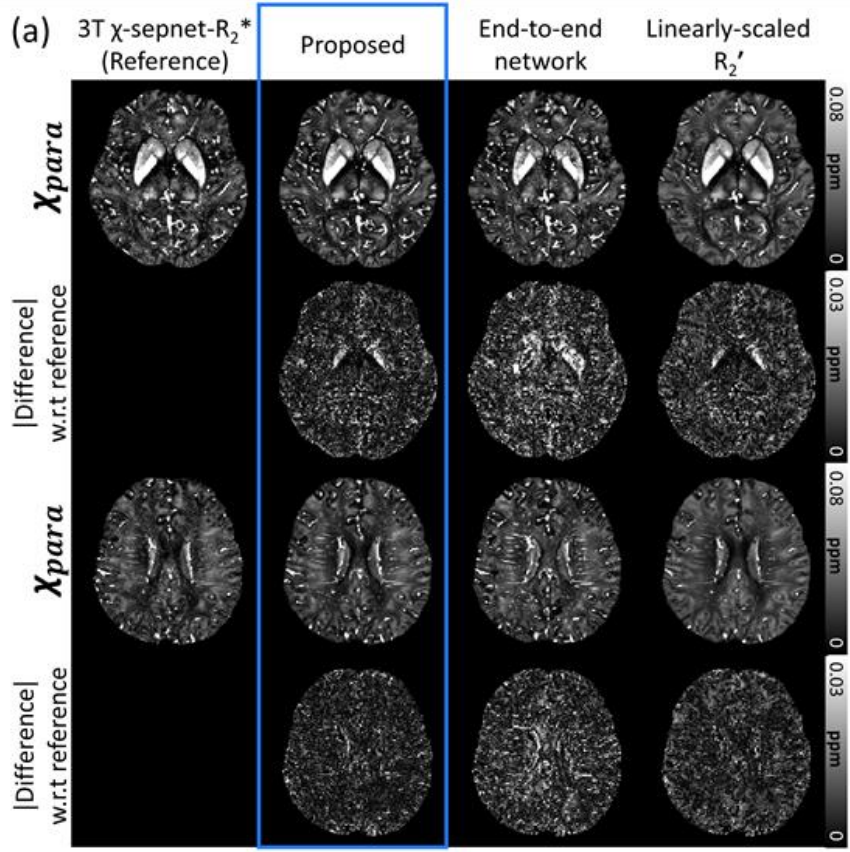
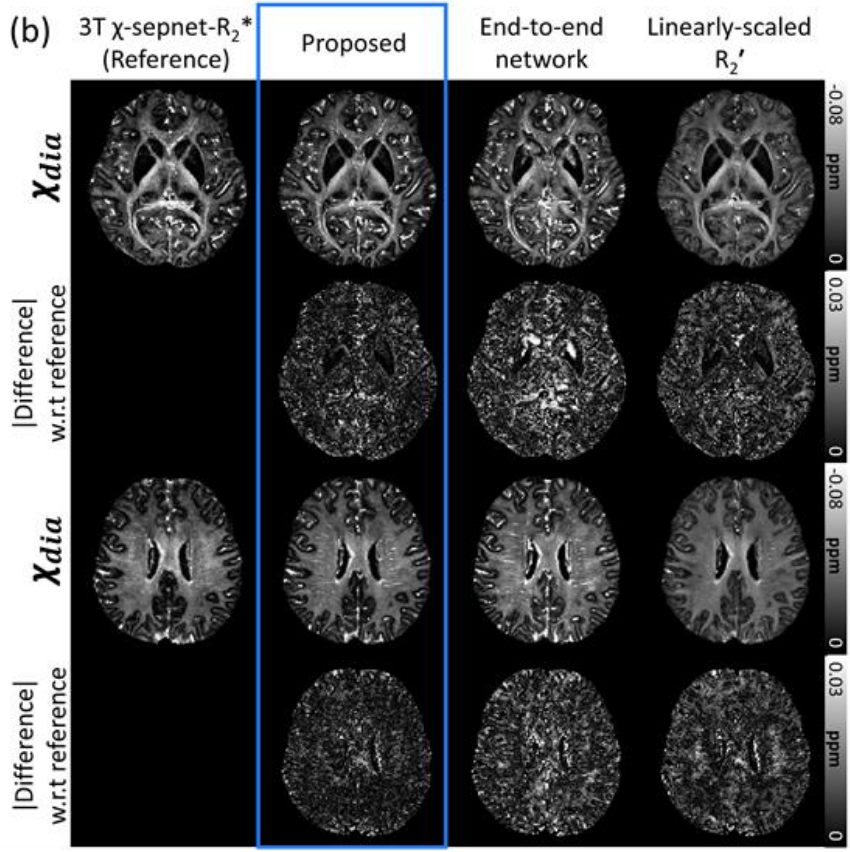

**Figure 3.** Comparison between χ-separation maps from χ-sepnet-$R_2^*$ at 3T (reference; first column), proposed method (blue boxes, second column), end-to-end network (alternative 1, third column), and linearly-scaled $R_2'$ (alternative 2, last column). All maps are at 1 mm isotropic resolution. (a) and (b) denote paramagnetic susceptibility maps ($\chi_{para}$) and diamagnetic susceptibility maps ($\chi_{dia}$), respectively. In the difference maps, CSF and vessels are masked out. The proposed method (blue boxes) demonstrates susceptibility contrasts closely matching to those of the reference.

**Table 3**. Evaluation of the three 7T χ-separation methods using pSNR, NRMSE, HFEN, and SSIM. All values are calculated with respect to the 3T χ-sepnet-$R_2^*$ maps. Higher pSNR, lower NRMSE, lower HFEN, and higher SSIM values indicate better correspondence to the reference maps.

|  |  | Proposed | End-to-end network | Linearly-scaled $R_2'$ |
|---|---|---|---|---|
| $\chi_{para}$ | pSNR (dB) | 39.7 ± 1.2† | 37.3 ± 0.8 | 39.3 ± 1.0 |
|  | NRMSE (%) | 42.1 ± 4.4† | 55.1 ± 4.3 | 43.9 ± 3.7 |
|  | HFEN (%) | 56.8 ± 7.6† | 76.4 ± 6.7 | 62.9 ± 6.3 |
|  | SSIM | 0.912 ± 0.031† | 0.878 ± 0.029 | 0.896 ± 0.037 |
| $\chi_{dia}$ | pSNR (dB) | 40.5 ± 1.1† | 38.5 ± 0.5 | 39.8 ± 0.7 |
|  | NRMSE (%) | 36.5 ± 2.3† | 46.1 ± 2.6 | 39.3 ± 1.3 |
|  | HFEN (%) | 56.2 ± 6.1† | 77.3 ± 3.9 | 65.6 ± 4.1 |
|  | SSIM | 0.925 ± 0.025† | 0.900 ± 0.030 | 0.907 ± 0.032 |

† Best metric among the three 7T χ-separation methods

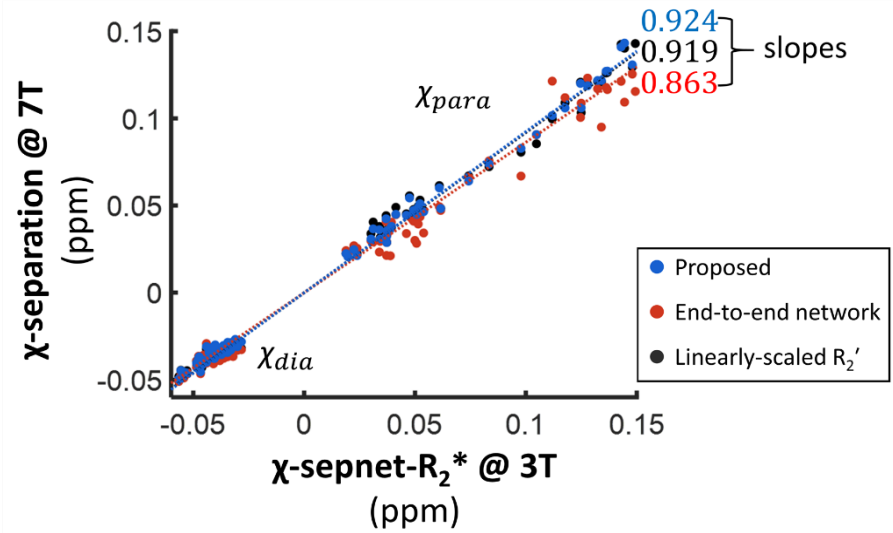

**Figure 4.** Results of the ROI analysis using linear regression. Each point represents the mean ROI value of a subject from 3T $\chi$-sepnet-$R_2^*$ on the x-axis and the corresponding value from the 7T $\chi$-separation method on the y-axis. The positive values are from the $\chi_{para}$ ROIs while the negative values are from the $\chi_{dia}$ ROIs. When the linear regression is conducted, the slopes are 0.924 (proposed), 0.863 (end-to-end network), and 0.919 (linearly-scaled $R_2'$). These results show the strongest consistency between the proposed method and 3T $\chi$-separation measurements.

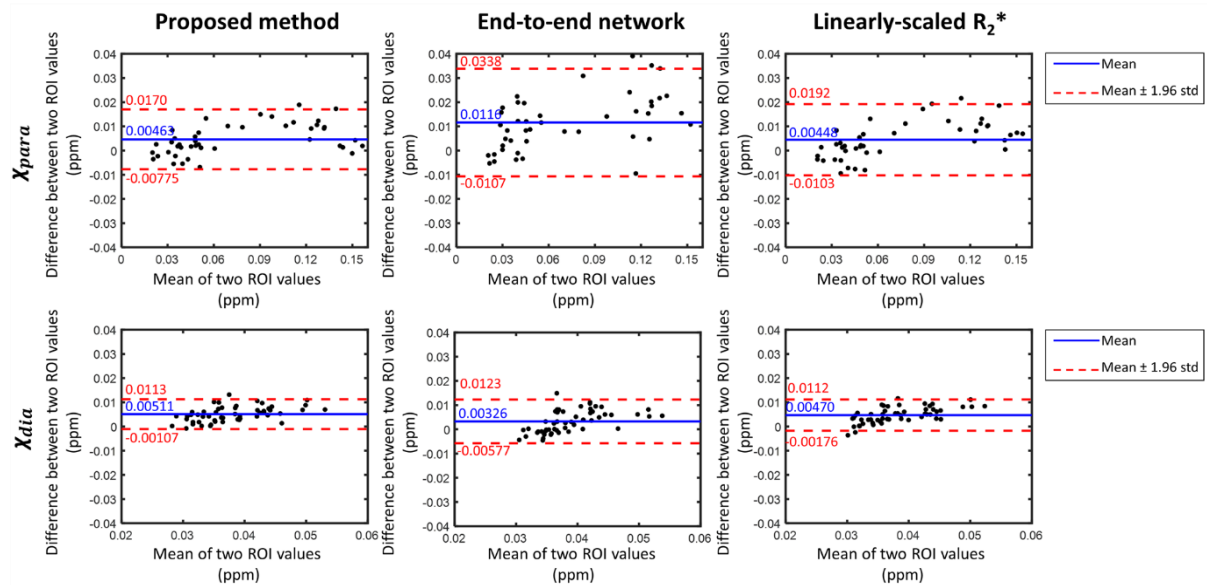

**Figure 5.** Results of the ROI analysis using Bland-Altman plots. The x-axis represents the mean of the two ROI values from a 7T $\chi$-separation method and 3T $\chi$-sepnet-$R_2^*$, while the y-axis indicates the difference between the two ROI mean values (3T $\chi$-sepnet-$R_2^*$ - 7T $\chi$-separation). The blue line represents the mean of the difference, and the red dashed line

indicates the mean ± 1.96 × standard deviation of the difference values. The proposed method shows the narrowest limits of agreement for both contrasts.

*3.3 Delineation of fine structures using high-resolution χ-separation at 7T*

In Figure 6, the χ-separation maps at 3T (first column for $\chi_{para}$ and third column for $\chi_{dia}$) with 1 mm isotropic resolution are compared to the χ-separation maps at 7T (second column for $\chi_{para}$ and fourth column for $\chi_{dia}$) obtained using the proposed method with 0.65 mm isotropic resolution. The overall contrasts presented at 3T and 7T are similar, although 7T maps tend to show slight underestimation (see Discussion).

Figure 7 demonstrates three zoomed-in areas of the 7T high-resolution χ-separation results that enable us to identify fine structures in the *in-vivo* human brain. In the motor cortex of the $\chi_{para}$ map (Fig. 7a), the hand knob region shows hyperintensity when compared to the corresponding area of the sensory cortex, consistent with previous findings in QSM (Stüber et al., 2014). In Fig. 4b, claustrum, which is a thin sheet of neurons wired to the coretex and subcortex, is outlined in the high-resolution $\chi_{para}$ map, agreeing with the χ-separation atlas at 3T (Min et al., 2024). Observing this type of tiny structures, which are not visible in most MRI images, may have important values. Lastly, transverse pontine fibers (yellow arrow) and fissures of cerebellum (orange arrow) are clearly visible in the $\chi_{dia}$ map (Fig. 7c), qualitatively agreeing with a previous study while providing a much more conspicuous contrast (Deistung et al., 2013a). These structures are less visible in the 1 mm 3T χ-separation maps of the corresponding locations.

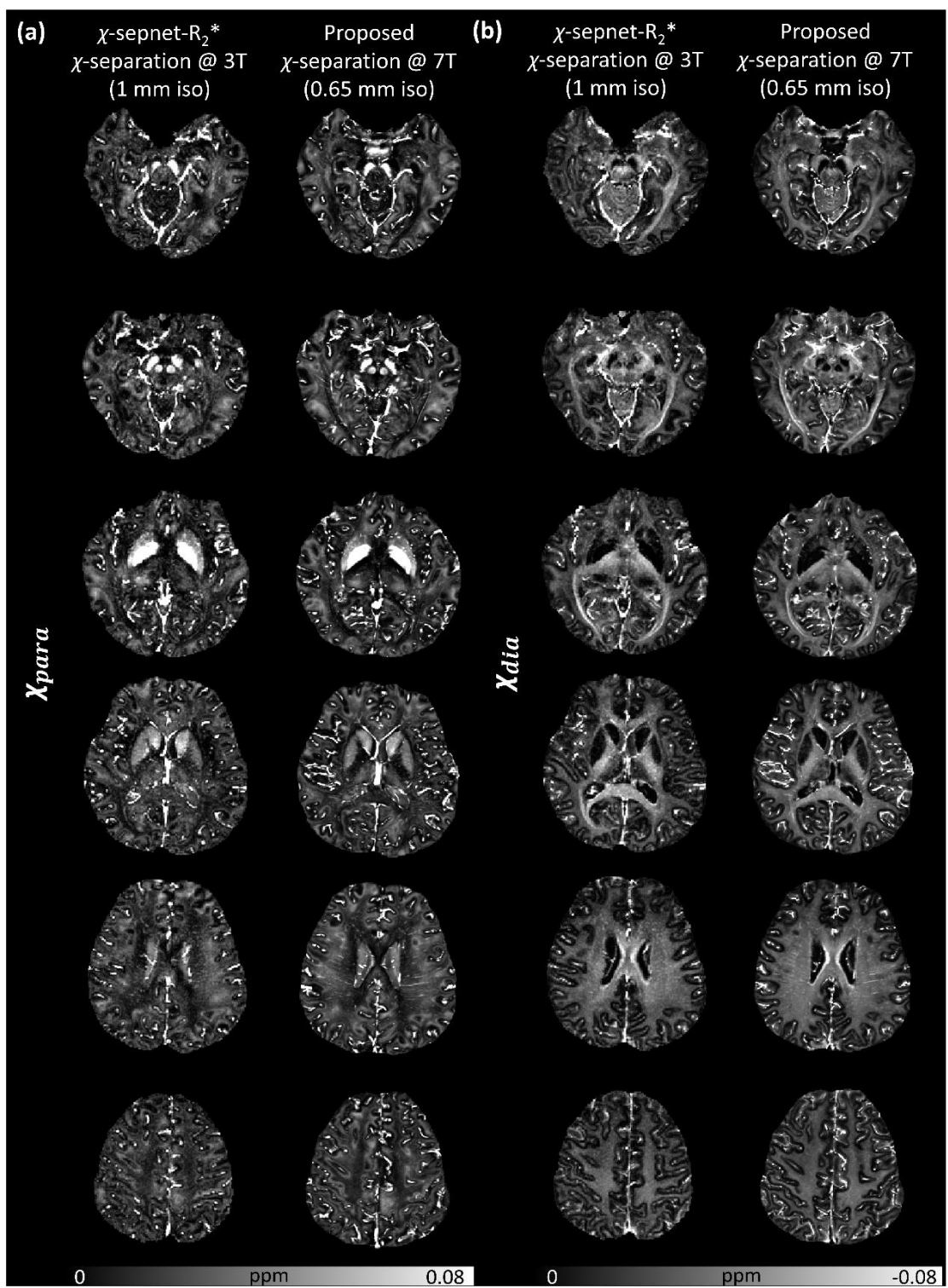

**Figure 6.** Comparison between 1 mm resolution χ-separation maps at 3T and 0.65 mm resolution χ-separation maps at 7T. The 7T maps reveal a clear and refined delineation of brain structures with less noise, while maintaining overall contrasts similar to those of the 3T maps.

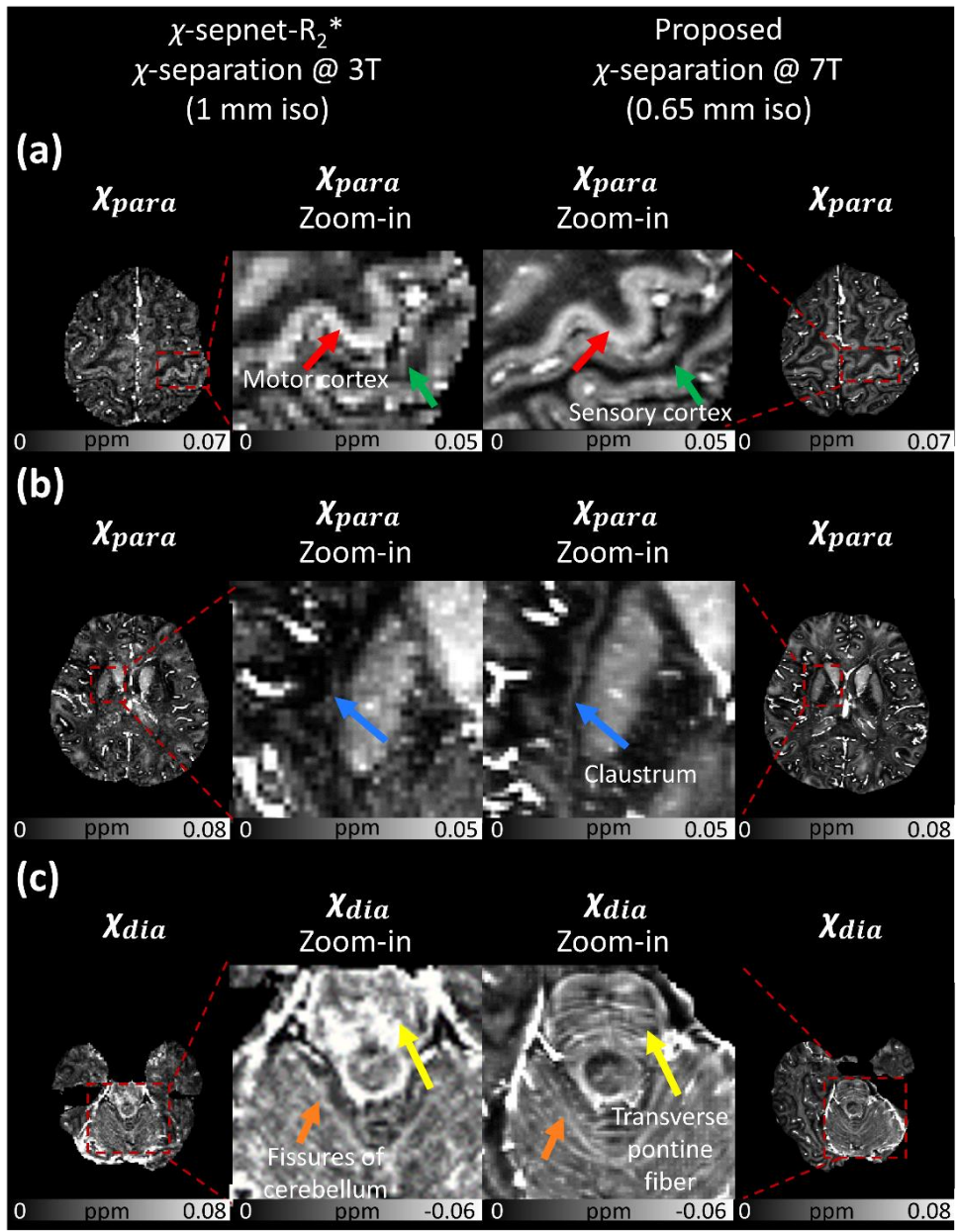

**Figure 7.** Delineation of fine structures using high-resolution χ-separation at 7T. Comparison is made between 1 mm isotropic resolution maps at 3T (left) and 0.65 mm isotropic resolution at 7T (right) at the slices of similar locations. (a) The hand knob area in the motor cortex (red arrow) and sensory cortex (green arrow). (b) Delineation of claustrum (blue arrow). (c) The $\chi_{dia}$ maps of pons and cerebellum, revealing transverse pontine fibers (yellow arrow) and fissures of cerebellum (orange arrow).

## 4. Discussion

This study presents a pipeline for *in-vivo* high-resolution χ-separation at 7T MRI. A key component of the pipeline is a novel deep neural network, R2PRIMEnet$_{7T}$, designed to eliminate $R_2$ acquisition, which is challenging at ultra-high field MRI. The results demonstrated that R2PRIMEnet$_{7T}$ can transform a 7T $R_2$* map to a 3T $R_2'$ map, successfully creating high-quality χ-separation results from multi-echo GRE data at 7T.

Application of the proposed high-resolution χ-separation method demonstrates that the high-resolution maps at 7T provide more refined details compared to χ-separation results from 3T, highlighting the advantage of 7T imaging. It is noteworthy that the acquisition time for 7T was only 8 minutes at the resolution of 0.65 mm, when covering a whole brain in a tight FOV. This protocol and the high-quality results highlight the potential of the proposed χ-separation at 7T as an effective tool for investigating neurodegenerative diseases, presurgical mapping, and layer-wise profiling of the cortex (Lee et al., 2023) at an individual level.

While deep learning methods, such as R2PRIMEnet$_{7T}$, χ-sepnet, and QSMnet$_{ME}$, offer promising results in our study, it is essential to acknowledge the limitations inherent to these approaches. One potential concern is error propagation when cascading multiple deep neural networks, particularly when faced with inaccuracies or artifacts in input data. This might manifest as inaccuracies in the $R_2'$ conversion process or $\chi_{para}$ and $\chi_{dia}$ maps reconstruction using χ-sepnet. In an attempt to evaluate this effect, we experimented with the end-to-end framework. However, the network exhibited worse performance than our proposed approach (Figure 3 and Table 3). We hypothesize that this could be due to the complexity of the task potentially requiring a larger network size, and issues with resolution dependency during the dipole deconvolution process as previously noted in QSMnet (Jung et al., 2022). Further research is necessary to explore the details of the effects of network cascading.

Due to the difficulties in acquiring multi-orientation GRE and $R_2$ data at 7T, we were unable to train the network using χ-separation maps at 7T as labels. Consequently, our current best approach was to use χ-sepnet trained on 3T χ-sep-COSMOS data or to train the network with 3T χ-sep-COSMOS data as the label.

Despite achieving favorable results in reconstructing high-resolution χ-separation maps with high quality and correspondence to results at 3T, the current focus is predominantly on data from healthy subjects and is restricted to specific resolutions (> 0.60 mm). To broaden the applicability of our proposed method, additional investigations should explore patients with altered susceptibility distribution and scan parameters such as anisotropic high-resolutions. For data with resolutions below 0.5 mm, a bottleneck may exist due to the limit of resolution generalization technique of QSMnet (Ji et al., 2023), making χ-separation of these data currently unfeasible. Addressing this limitation in future research is important and necessary.

When calculating voxel-wise metrics such as pSNR, NRMSE, HFEN, and SSIM, registration was performed on the 7T maps to match the 3T reference maps. This registration process may have introduced errors in the metrics. Furthermore, the presence of noise in the reference maps at 3T may have also influenced the metrics, such as NRMSE and SSIM, particularly in the $R_2'$ maps (Wang et al., 2024). Therefore, these aspects should be taken into consideration when interpreting the results.

Despite the effectiveness of the proposed method in visualizing susceptibility-related brain structures, both 3T and 7T χ-separation maps exhibit common artifacts, such as high values in vessels. These artifacts are well-recognized in previous studies (Lee et al., 2024) and arise due to flow and non-local effects induced by large vessels. In one attempt to address this, we masked out vessels during voxel-wise analysis. However, intrinsic issues persist and should be mitigated in future work.

As demonstrated by the results of the ROI analysis (Figures 4 and 5) and Figure 6, the proposed method exhibits slight underestimation, evidenced by a linear regression slope lower than 1.0 and lower contrast. We hypothesize that this discrepancy may arise from the underestimation of QSM values at 7T (Figure S6). This underestimation of QSM could be attributed to the impact of field strength on voxel size discretization and magnetic susceptibility (Chen et al., 2022; Rua et al., 2020; Spincemaille et al., 2020). These results suggest that there are limitations in using the 3T results as a reference, as discussed in detail in the following paragraph.

In overcoming challenges inherent in χ-separation at 7T, including the requisite for high-resolution $R_2$ mapping, we employed χ-sepnet and R2PRIMEnet$_{7T}$. Our results demonstrate the concordance with χ-separation at 3T, a method confirmed for its accuracy, efficacy and utility through previous studies (M. Kim et al., 2024; Kim et al., 2022; Lee et al., 2023; Min et al., 2024; Müller et al., 2024; Rovira and Pareto, 2024; Zhu et al., 2024). However, further validation of our study's accuracy hinges on the acquisition of the whole-brain $R_2$ map at 7T for χ-separation. To achieve this, further research is necessary, including the development of high-resolution $R_2$ mapping techniques. Additionally, if an $R_2$ map is successfully acquired, another hurdle arises in selecting a relaxometry constant value ($D_r$), a linear constant between $R_2'$ and susceptibility, because it varies depending on diffusion, susceptibility source characteristics, and B0 field strength (Carr and Purcell, 1954; Emmerich et al., 2021a; Shin et al., 2024; Yablonskiy and Haacke, 1994). Lastly, refining the model to enhance the accuracy of χ-separation at 3T may positively impact the improvement of our proposed method at 7T (Kan et al., 2024; Li et al., 2023).

## 5. Conclusion

In conclusion, our study presents a novel methodology for *in-vivo* high-resolution χ-separation at 7T, addressing challenges associated with $R_2$ mapping at 7T. The results showcase a good agreement with the χ-separation results at 3T, surpassing the performance of the alternative methods. Importantly, our proposed method opens avenues for enhanced delineation of intricate brain structures as demonstrated in our 0.65 mm isotropic resolution results. This work contributes to *in-vivo* high-resolution susceptibility mapping, providing a quantitative tool for researchers and clinicians exploring brain architectures related to iron and myelin distribution.